# Resonance effects in photoemission time delays

M. Sabbar[1*], S. Heuser[1], R. Boge[1], M. Lucchini[1], T. Carette[3,4], E. Lindroth[4], L. Gallmann[1,2], C. Cirelli[1#] and U. Keller[1]

[1]Physics Department, ETH Zurich, 8093 Zurich, Switzerland
[2]Institute of Applied Physics, University of Bern, 3012 Bern, Switzerland
[3]Laboratoire de Chimie Quantique et Photophysique, CP160/09, Université Libre de Bruxelles, B 1050, Brussels, Belgium
[4]Physics Department, Stockholm University, AlbaNova University Center, SE-106 91, Stockholm

[*]msabbar@phys.ethz.ch
[#]cirelli@phys.ethz.ch

**Abstract:** We present measurements of single-photon ionization time delays between valence electrons of argon and neon using a coincidence detection technique that allows for the simultaneous measurement of both species under identical conditions. Taking into account the chirp of the ionizing single attosecond pulse (attochirp) ensures that the 'clock' of our measurement technique is started at the same time for both types of electrons, revealing with high accuracy and resolution energy-dependent time delays of a few tens of attoseconds. By comparing our results with theoretical predictions, we confirm that the so-called Wigner delay correctly describes single-photon ionization delays as long as atomic resonances can be neglected. Our data, however, also reveal that such resonances can greatly affect the measured delays beyond the simple Wigner picture.

Recent measurements have demonstrated the possibility of probing single-photon ionization time delays of electrons originating from different initial states [1, 2]. The controversy about the nature and interpretation of these time delays triggered many theoretical efforts [3-8] revealing the measured time delay to be composed of two different contributions: a measurement induced delay which can be subtracted using computational results and the actual atom-specific ionization delay identified as the Wigner time delay [9, 10]. Our coincidence measurements presented here provide an unsurpassed precision and confirm the general trend of the Wigner time delays for single-photon ionization for two different atomic species argon (Ar) and neon (Ne) over an extreme ultraviolet (XUV) photon energy range of 28 to 40 eV, but also directly reveal the significant influence of atomic resonances beyond the simple Wigner picture for the first time.

The Wigner time delay is defined as a measure for the spectral variation of the scattering phase. Scattering theories apply well to the mechanism of single-photon ionization because this process can be seen as a half-scattering event: after the electron is promoted from a bound state into the continuum, the electron scatters off the attractive Coulomb potential of the ion. The Wigner time delay $\tau_W$ is calculated from the phase difference $\varphi_W$ between an electron wavepacket propagating through the potential and that of a free particle. This particular delay definition represents the group delay of the wavepacket referenced to the motion of the free particle with the same kinetic energy: $\tau_W \doteq \partial \varphi_W / \partial \omega = \hbar \partial \varphi_W / \partial E$. Because of this free-particle reference, the Wigner delay only considers delays that stem purely from the interaction of the electron with the potential.

Our earlier experiments on ionization time delays in the tunneling regime have shown that the Wigner delay is not always a useful concept [11, 12]. Following the peak of the wavepacket with the group delay (or Wigner delay) for tunneling is particularly tricky, because first, it is not clear when the tunneling should exactly start and second, the energy-dependent transmission will reshape the wavepacket such that the peak has no meaning for the tunneling time. In contrast to a light pulse, an electron wavepacket disperses even in vacuum. Since the propagation of the peak of the wavepacket is defined by the group delay, almost any group delay can be measured during propagation in combination with an appropriate energy-dependent transmission filter. The main difference between single-photon and tunnel ionization can be explained with a simplified picture of a wavepacket propagation through a square potential exploring different regimes of $E<E_b$ for the tunnel ionization and $E>E_b$ for the single-photon ionization (Fig. 1). The tunnel barrier exhibits an energy dependent high-pass transmission filter (Fig. 1(a)). Figure 1(b) represents the case of a wavepacket before and after the potential barrier with a kinetic energy $E<E_b$. Thus for energies $E<E_b$ transmission is much more likely on the high-energy side of the wavepacket. Through its selective transmission, this filter considerably reshapes the wavepacket. This leads to the formation of a new peak of the wavepacket that will not correctly describe the tunneling time as recenty shown experimentally with the attoclock technique [11, 12].

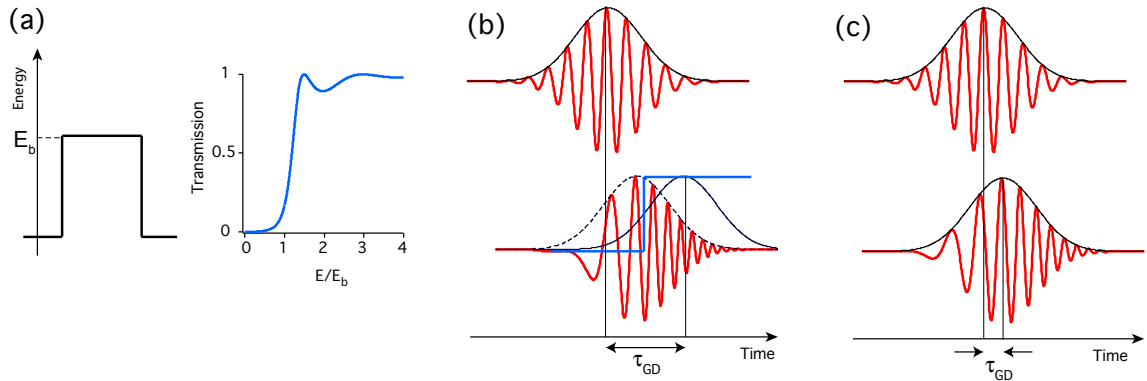

FIG. 1: Propagation of a wavepacket through a square potential barrier. (a) Potential barrier of height $E_b$ (in black) and transmission probability (in blue) as a function of the ratio between the kinetic energy of the wavepactet E and the barrier height $E_b$. (b) If the average kinetic energy of the wavepacket is smaller than the barrier height, propagation involves tunneling. After propagation through the barrier, the wavepacket disperses and its peak would be found at a time given by the dashed line; however, the transmission probability, shown in (a) acts as an energy-dependent high-pass filter (in blue), inducing an additional temporal shift, which is not related to the time spent in the barrier but rather only depends on the specific shapes of filter and wavepacket. (c) If the kinetic energy of the wavepacket is larger than $E_b$, the group delay is a direct measure of the propagation time of the wavepacket because the energy-dependence of the filter can be neglected in that regime and a meaningful relation between the wavepacket peaks before and after the filter persists.

On the other hand, if the electron wavepacket propagates with a kinetic energy sufficiently larger than the barrier height (Fig. 1(c)), the energy-dependence of the amplitude filter can be neglected because the transmission probability is close to one for all the energies within the bandwidth of the wavepacket (the modulations after the point $E=E_b$ shown in Fig. 1a are due to the reflected waves at the well boundary). In contrast to the tunneling regime, the Wigner delay is therefore expected to be a good concept to estimate single-photon ionization time delays, as has already been proposed theoretically [3, 4, 13] and demonstrated experimentally [2, 14, 15].

Here we show that atomic resonances act as an additional energy filter and therefore reshape the wavepacket, which leads to the breakdown of the simple scattering group delay picture and to significant deviations from a spectrally feature-less Wigner delay. We present experimental results that allow the determination of the photoionization delay difference between the valence electrons of Ar (3p) and Ne (2p). The novel experimental scheme applied here combines the attosecond streaking technique [16] with coincidence detection [17, 18]. The unique ability to assign electrons to their parent ions allows to simultaneously record multiple photoelectron spectra originating from different species even when the kinetic energies overlap. This capability and the careful treatment of the chirp of the attosecond pulse

(attochirp) [19], allows us to extract the one-photon delay difference between Ar and Ne wavepackets. Our measurements show a general trend with calculated Wigner delays for energies between 28-35 eV, however, we also observe strong deviations of the measured data from the Wigner delay predicted by single-channel scattering theory which can be explained by the presence of resonances in Ar [20].

The technique used to conduct the experiment is based on a reaction microscope, also known as a cold target recoil ion momentum spectroscopy (COLTRIMS) detector [17, 18], in combination with a gas target containing a mixture of both species [21]. The gas mixture is ionized with single attosecond pulses (SAPs) of 12 eV bandwidth centered at a photon energy of about 35 eV. A moderately strong (about $3 \times 10^{12}$ W/cm$^2$) infrared (IR) pulse with controllable delay modulates the final photoelectron momenta by streaking the freed electrons. The SAPs are generated with the polarization gating technique [22, 23] using waveform controlled few-cycle IR laser pulses at a center wavelength of 735 nm and with a pulse duration of approximately 6 fs focused into an Ar gas target. The XUV-pump beam is first recombined with the delayed IR-probe through a holey mirror. Both beams are then collinearly focused by a toroidal mirror into the COLTRIMS detector, where ions and electrons are separated by a uniform DC electric field and guided towards space and time sensitive detectors. This allows for retrieving the full momentum vector - and therefore the kinetic energy - of each individual particle at the moment of ionization. Thus, applying a filter to the time-of-flight of the parent ions and to the momentum sum of ions and electrons allows for coincidence detection.

Based on attosecond streaking in coincidence we have been able to distinguish between electrons generated from Ar and Ne even though they energetically overlap. Figure 2(a) shows the streaking traces simultaneously recorded for each species. For our analysis, we only consider electrons emitted into a cone with an opening angle of 20 degrees with respect to the XUV polarization axis. In order to retrieve the energy dependent phase of the photoelectron wavepacket, an algorithm known as frequency-resolved optical gating for complete reconstruction of attosecond bursts (FROG-CRAB) [24, 25] has been employed.

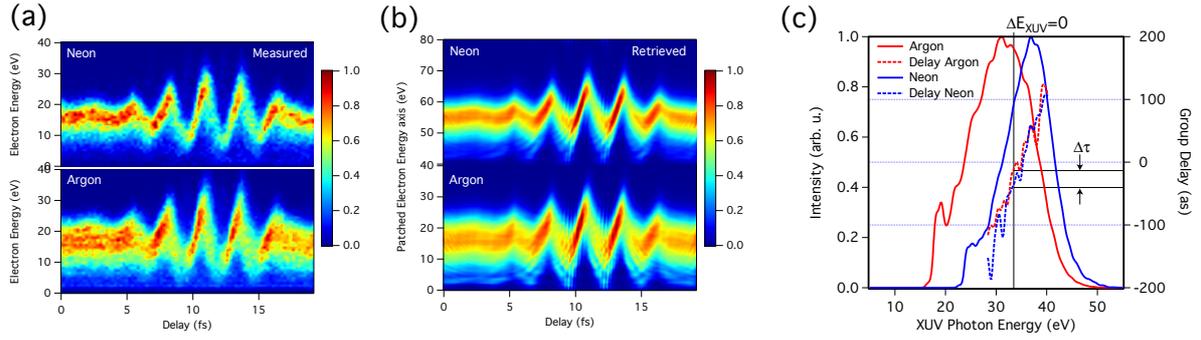

FIG. 2. Measured (a) and reconstructed (b) spectrograms for Ne and Ar photoelectrons; before applying the retrieval algorithm the two traces are patched together onto a common energy axis in order to ensure consistency in the reconstruction (see main text). (c) XUV spectra (solid lines) and group delays (dashed lines) for Ar (red) and Ne (blue) computed by adding the ionization energy of the two targets to the spectra and group delays of the photoelectron wavepackets retrieved with the FROG-CRAB algorithm. The vertical and horizontal black lines indicate the resulting group delay difference calculated at the same XUV photon energy: this procedure ensures that the attochirp contribution to the photoemission time delay is removed.

The algorithm has been fed with a matrix where both, the Ar and the Ne trace, had been patched together as illustrated in Fig. 2. This procedure makes sure that the same IR vector potential and the same time zero are used for the reconstruction of the electron phase of both Ar and Ne.

The resulting spectra and group delays are shown in Fig. 2(c). For both target atoms the group delay curve exhibits a large slope, on the order of 25 as/eV, indicating that the XUV pulse has a relatively strong chirp. This means that XUV photons of different energy ionize the target atoms at different times. If we compared the group delays for Ar and Ne electron wavepackets at the same kinetic energy, an apparent delay would arise simply because the electrons have started at different instances of time. However, we can simply cancel the attochirp contribution to the photoemission time delay evaluating the group delay difference between Ar and Ne, $\Delta\tau^{Ar/Ne}$, at the same XUV photon energy as shown in Fig. 2(c).

The difference between the group delays, calculated for any XUV energy within a range where the spectral intensity of Ar and Ne spectra overlaps (between 28 and 40 eV) results in the energy-dependent group delay curve presented in Fig. 3(a) as green open circles. The data represent the averaged delays of 33 independently measured traces, while the error bars represent the standard deviation between the different data sets.

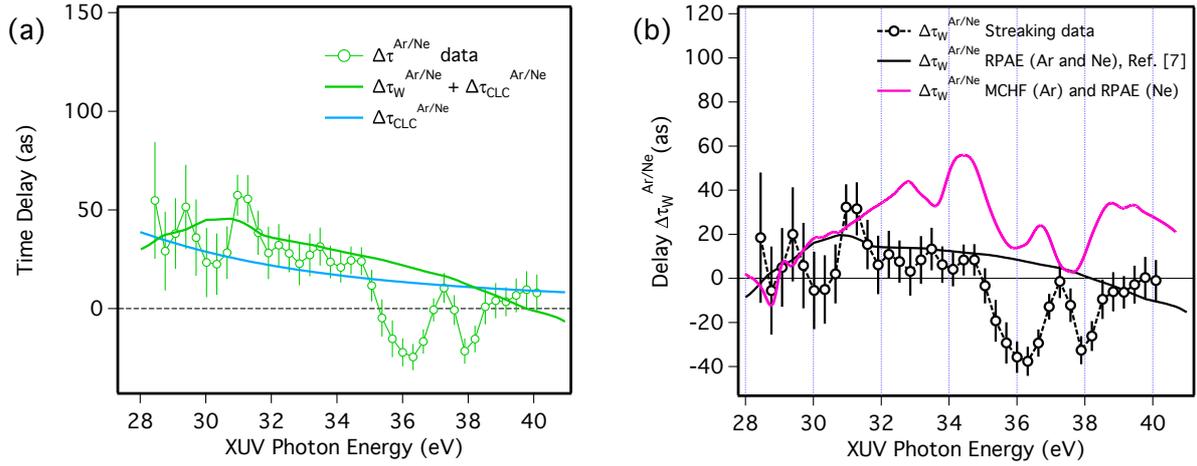

FIG. 3. (a) The measured streaking group delay difference between Ar and Ne electrons (green open circles) is compared with the theoretical prediction (green solid line). The green solid line represents the sum of the calculated Wigner delay difference of Ar and Ne [7] and the corresponding measurement-induced delay difference [26], shown as a light blue curve. (b) The black open circles represent the averaged group delay difference of the 33 independent measurements obtained after subtracting the laser-induced contribution shown in (a). The black and magenta solid lines represent theory curves. The black line is obtained by computing the one-photon matrix elements within the RPAE [7] both for Ar and Ne electrons. The magenta line has been obtained by using MCHF [20] for Ar taking into account resonances and RPAE [7] for Ne.

Recent advancements in computational methods for solving the time-dependent Schrödinger equation (TDSE) [4, 26] have allowed for a more accurate description of the time-dependent photoionization process. Solving the TDSE in the single-active electron approximation demonstrated that the delay extracted from attosecond streaking experiments is identical with the Wigner time delay $\tau_W$ only in the case of a short-range model potential (for instance, like a Yukawa potential) [4]. More generally, the presence of a Coulomb potential introduces an additive time delay, $\tau_{CLC}$ (Coulomb-laser coupling), which originates from the interaction of the streaking field with the long-range asymptotic tail of the Coulomb potential:

$$\tau_{streaking} = \tau_W + \tau_{CLC}.$$

The measurement-induced contribution $\tau_{CLC}$ can be extracted from numerical calculations by computing the difference between the delays determined for a short-range and a Coulomb potential. It has been shown that this contribution is, to a great extent, universal, depending only on the net ion charge, the final electron energy and the central frequency of the streaking field [26]. It is worth emphasizing that, to leading order, the calculated $\tau_{CLC}$ has no dependence on the IR intensity [4].

In order to compare our experimental results to computational results, in a first approach we have taken the calculated Wigner delay for Ar and Ne from Ref. [7] which uses the random-phase approximation with exchange (RPAE) and thereby takes care of many-electron correlation. In our analysis we considered only the 3p→Ed channel for Ar and the 2p→Ed channel for Ne, respectively, because ionization from 3s (for Ar) or 2s (for Ne) shell is supposed to be much weaker in the photon energy range considered in this work.

The $\tau_{CLC}$ part has been taken from Ref. [26]. Subtracting the laser-induced contribution from the experimental data presented as green circles in Fig. 3(a), we obtain the difference between the Wigner delays of Ar and Ne as shown in Fig. 3(b) (black open circles). In the energy region between 28 and 35 eV, our data confirm the calculated Wigner time delays of Ref. [7] within the accuracy of the experiment.

However, in the energy range between 35 and 39 eV the deviation of the data from the theory is substantial, showing evidence of the presence of additional delays beyond the scattering picture provided by the Wigner delay. As we will show, a novel *ab initio* method, based on a multi-configurational Hartree–Fock (MCHF) [20] strongly suggests that these sharp features are due to multiple resonances originating from shake-up thresholds in Ar opening within this energy range. Like RPAE, MCHF includes contributions from different angular channels, but also accounts for the influence of doubly excited states and ionization thresholds that lead to multiple resonance structures. It is known that the $Ar^+$ level structure is particularly rich in the energy range between 35 and 39 eV and first theoretical evidence has been brought that the presence of resonances decaying into $Ar^+$ ($3s^23p5$) and $Ar^+$ ($3s3p^6$) may greatly affect the measured one-photon delay [20, 27]. We calculated the one-photon Wigner delay for Ar with MCHF for an outgoing d-wave and subtract the one-photon Wigner delay for Ne estimated with RPAE. The latter is expected to be accurate due to the absence of resonances in the $Ne^+$ spectrum in the considered energy range. The Ar atomic structure model used here is described in more details in the Appendix A of Ref. [20]. The Wigner delays computed with this model are obtained from photoelectron amplitudes convoluted by a Gaussian IR pulse with a full width at half maximum (FWHM) of 0.4 eV.

The result of this computation is shown in Fig. 3(b) as a magenta line. Its discrepancy to the experimental data can partly be attributed to an inaccurate description of the 3p ionization Cooper minimum in Argon. Indeed, the Cooper minimum at about 50 eV accelerates 3p photoelectrons by several tens of attoseconds in a broad range of energies, including the range of interest [28]. The Cooper minimum in the MCHF simulations is too high and narrow, which leads to overestimations of the delays below 40 eV as large as up to 15 as.

More relevant to our discussion is the qualitative agreement between theory and the experimental data in the energy range between 35 and 39 eV which demonstrates the relevance of the atomic resonances for the determination of photoionization time delays. Note that the structure in the recorded photoemission delays appears about one IR photon above the two groups of resonances that affect the most the calculated cross-sections.

Photoionization in the region from 33.5 eV to 35 eV exhibits a large number of resonances [29]. According to the calculations, the first group of resonances results from two $3p^4nl\ ^2D^e$ thresholds, at about 34.3 eV, and the second results from a $^2S^e$ threshold, at 36.5 eV. Future theoretical investigations are expected to confirm the attribution of the structure seen in the experimental time delays, since the presence of resonances can affect measured delays beyond their effect on the single-photon spectrum [27] which has been applied here. Furthermore, a proper treatment of s electrons and the experimental angular integration might impact the results, particularly close to resonances where the relative channel amplitudes change rapidly.

In conclusion, we have accessed photoionization time delays between valence electrons of two different atomic species by taking advantage of the unique capabilities of a COLTRIMS detector and the attosecond streaking technique. The time delays retrieved by our measurements confirm the general trend of scattering theories based on the Wigner delay picture for a photon energy range where resonances in Ar are not present. However, our data also clearly reveal the influence of these resonances on photoionization time delays beyond the simple Wigner picture.

The authors would like to thank M. Dahlström, R. Pazourek, and S. Nagele for fruitful discussions. This work was supported by the ERC advanced grant ERC-2012-ADG_20120216 within the seventh framework program of the European Union and by the NCCR MUST, funded by the Swiss National Science Foundation. M. L. acknowledges support from the ETH Zurich Postdoctoral Fellowship Program. T. C. acknowledges support from the IUAP - Belgian State Science Policy (BriX network P7/12). E. L. and. acknowledge support from the Swedish research council (VR).